\documentclass[%
 reprint,
 amsmath,amssymb,
 aps,
]{revtex4-1}
\usepackage{graphicx}
\usepackage{dcolumn}
\usepackage{bm}

\begin{document}

\title{Strain effects on Phase-Filling Singularities in Highly Doped $n$-Type Ge}%


\author{Zhigang Song}
\affiliation{School of Electrical and Electronic Engineering, Nanyang Technological University, 50 Nanyang Avenue, Singapore 639798, Singapore}
\author{Wei-Jun Fan}
\email{ewjfan@ntu.edu.sg}
\affiliation{School of Electrical and Electronic Engineering, Nanyang Technological University, 50 Nanyang Avenue, Singapore 639798, Singapore}
\author{C. S. Tan}
\affiliation{School of Electrical and Electronic Engineering, Nanyang Technological University, 50 Nanyang Avenue, Singapore 639798, Singapore}
\author{Qijie Wang}
\affiliation{School of Electrical and Electronic Engineering, Nanyang Technological University, 50 Nanyang Avenue, Singapore 639798, Singapore}
\author{Donguk Nam}
\affiliation{School of Electrical and Electronic Engineering, Nanyang Technological University, 50 Nanyang Avenue, Singapore 639798, Singapore}
\author{D. H. Zhang}
\affiliation{School of Electrical and Electronic Engineering, Nanyang Technological University, 50 Nanyang Avenue, Singapore 639798, Singapore}
\author{Greg Sun}
\affiliation{$^{2}$Department of Engineering,
University of Massachusetts Boston, Massachusetts 02125, U.S.A}

\date{\today}

\begin{abstract}
Recently, Chi Xu et al. predicted the phase-filling singularities (PFS) in the optical dielectric function (ODF) of the highly doped $n$-type Ge and confirmed in experiment the PFS associated $E_{1}+\Delta_{1}$ transition by advanced \textit{in situ} doping technology [Phys. Rev. Lett. 118, 267402 (2017)], but the strong overlap between $E_{1}$ and $E_{1}+\Delta_{1}$ optical transitions made the PFS associated $E_{1}$ transition that occurs at the high doping concentration unobservable in their measurement. In this work, we investigate the PFS of the highly doped n-type Ge in the presence of the uniaxial and biaxial tensile strain along [100], [110] and [111] crystal orientation. Compared with the relaxed bulk Ge, the tensile strain along [100] increases the energy separation between the $E_{1}$ and $E_{1}+\Delta_{1}$ transition, making it possible to reveal the PFS associated $E_{1}$ transition in optical measurement. Besides, the application of tensile strain along [110] and [111] offers the possibility of lowering the required doping concentration for the PFS to be observed, resulting in new additional features associated with $E_{1}+\Delta_{1}$ transition at inequivalent $L$-valleys. These theoretical predications with more distinguishable optical transition features in the presence of the uniaxial and biaxial tensile strain can be more conveniently observed in experiment, providing new insights into the excited states in heavily doped semiconductors.

\end{abstract}

\maketitle

\begin{figure}[htb]
\includegraphics[width=3.4in]{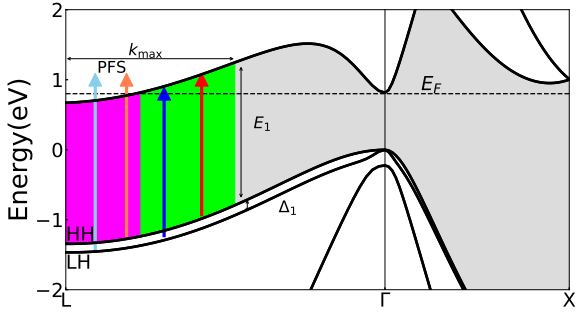}
\caption{Band structure of Ge calculated with the 30-band $k \cdot p$ model \cite{rideau2006strained,Zhigang,8871170}. Fermi level $E_{F}$ (dotted line) is raised 0.13 eV above the CBM at $L$-valley. The $E_{1}$ (red arrow) and $E_{1}+\Delta_{1}$ (blue arrow) optical transitions are allowed in the green region, but forbidden in the purple region due to the Pauli blocking. Orange and light-blue arrows in the purple region indicate the PFS associated $E_{1}$ and $E_{1}+\Delta_{1}$ optical transitions to higher CB states above the Fermi level.}
\label{fig1}
\end{figure}

Optical transitions between the conduction band (CB) and valence band (VB) in semiconductor are directly related with the van Hove critical point singularities \cite{peter2010fundamentals} in joint density of states (DOS), which can be observed from the sharp features in the complex optical dielectric function (ODF), $\varepsilon(E)$  that can be probed by the spectroscopic ellipsometers. For most direct band gap semiconductors in III-V and II-VI compounds, heavy doping suppresses suppresses the interband optical transition involving states between CB and VB edges due to the Pauli blocking effect. This leads to the Burstein-Moss shift \cite{Burstein1,Moss_1954} in the optical transition energy, which can be deduced from $\varepsilon(E)$. It has been theoretically predicted that the profile of complex ODF is modified in heavily doped semiconductors by the phase-filling singularities (PFS) as a result of the abrupt change in the occupation number at the Fermi level \cite{PhysRevLett.25.1486}.
Recently, the PFS was successfully observed in heavily doped $n$-type Ge \cite{Xuchi1} by the advanced low-temperature \textit{in situ} doping techniques with the additional PFS involving the so-called $E_{1}+\Delta_{1}$ optical transitions in the real and imaginary part of the optical dielectric function's second differential (ODFSD).
Fig. \ref{fig1} illustrates the $E_{1}$ and $E_{1}+\Delta_{1}$ optical transitions in the heavily doped n-type Ge with its Fermi level sitting above the $L$-valley CB minimum (CBM). As a result, the ordinary $E_{1}$ and $E_{1}+\Delta_{1}$ optical transitions are only allowed in the green region where the dispersions of CB and VB of both heavy hole (HH) and light hole (LH) run parallel, but not in the purple region due to the Pauli blocking effect. What are allowed in the purple region are the PFS associated $E_{1}$ and $E_{1}+\Delta_{1}$ optical transitions that take place to CB states sitting above the Fermi level with higher transition energies relative to their respective ordinary optical transitions.
However, the singularity feature originated from the $E_{1}$ optical transition was not observed in the heavily doped Ge because of the strong overlap between the $E_{1}$ and $E_{1}+\Delta_{1}$ optical transitions, only that associated with the $E_{1}+\Delta_{1}$ optical transition is observable.
An important question to answer is then what mechanism we can use to effectively separate the PFS between $E_{1}$ and $E_{1}+\Delta_{1}$ optical transitions, which would then allow for experimental observation of PFS-associated $E_{1}$ optical transition. Another practical consideration is whether we can lower the doping concentration required for the observation of PFS.

\begin{figure}[htb]
\includegraphics[width=3.4in]{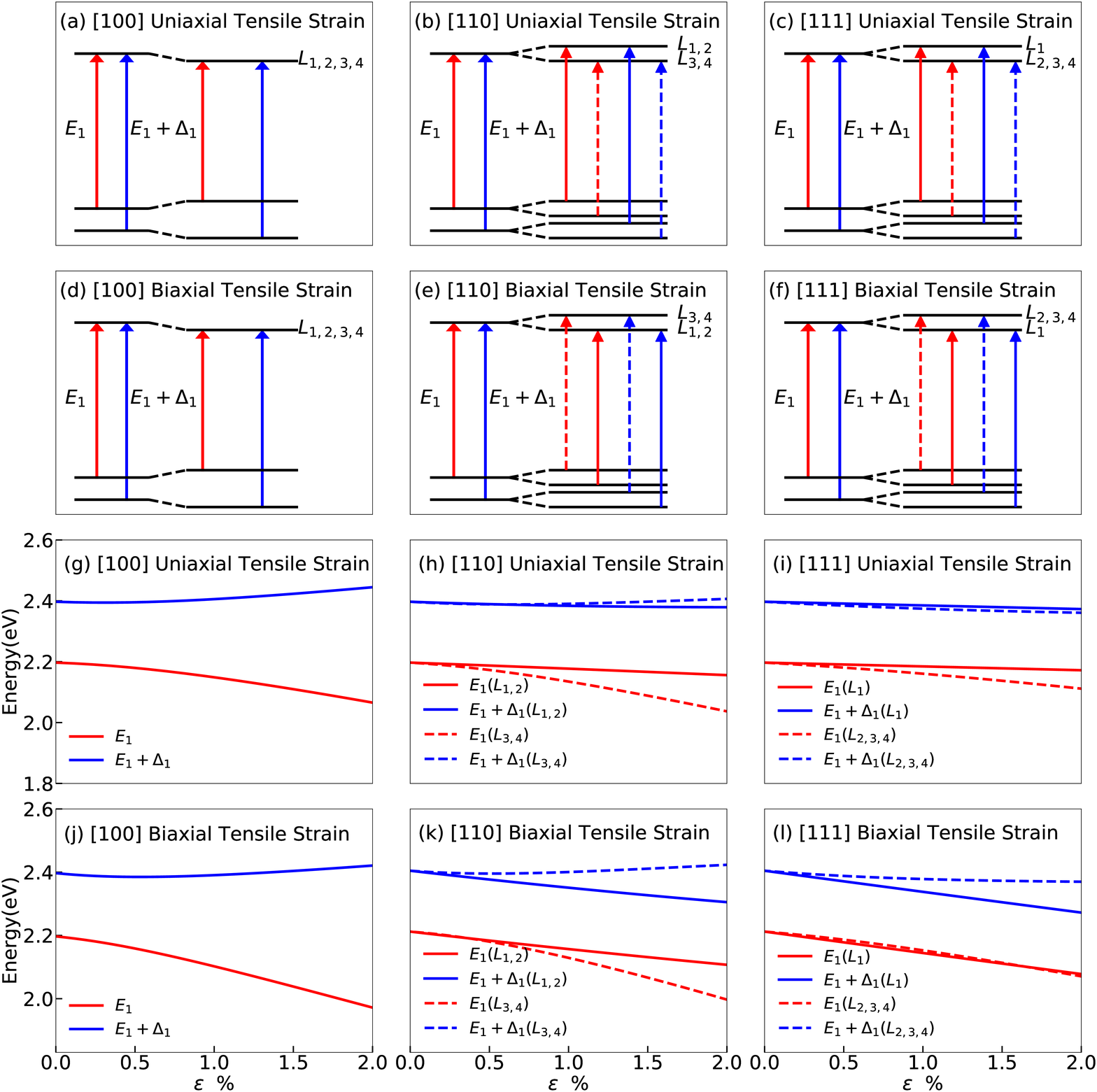}
\caption{(a)-(f) illustrate the $E_{1}$ and $E_{1}+\Delta_{1}$ optical transitions at the four $L$-valleys under either the uniaxial or biaxial strain applied along [100], [110] and [111] and (g)-(l) are the variation of those transition energies in the presence of the uniaxial or biaxial tensile strain along the different directions. The red curves and arrows denote the $E_{1}$ optical transitions and blue curves and arrows represent $E_{1}+\Delta_{1}$ optical transitions. The tensile strain along [100] shifts the $E_{1}$ and $E_{1}+\Delta_{1}$ transition energy by lowering the CBM and increasing $\Delta_{1}$ at $L$-valley as shown in (a) and (d), effectively widening the separation between $E_{1}$ and $E_{1}+\Delta_{1}$ transitions. The tensile strain along [110] and [111], on the other hand, break the degeneracy of the four CB $L$-valleys in Ge, splitting each of the $E_{1}$ and $E_{1}+\Delta_{1}$ optical transitions into two as shown by the solid and dashed arrows in (b) (c) (e) and (f).}
\label{fig2}
\end{figure}

The answer can be found in strain, as an essential tuning mechanism in band engineering, applied to the doped semiconductors. Since the PFS associated with the $E_{1}$ and $E_{1}+\Delta_{1}$ optical transitions originate from the partially suppression due to the parallel band, the most effective approach is therefore to make sure that the electrons from the $n$-type doping mostly enter the $L$-valley in the presence of strain. The occupiable number of electrons at given CB valley can be determined by the DOS effective mass $m_{\rm{DOS}}^{*}$ associated with the valley.
For Ge, $m_{L_{\rm{DOS}}}^{*}$ at $L$-valley is approximately the same as $m_{X_{\rm{DOS}}}^{*}$ at $X$-valley, both are much larger than $m_{\Gamma_{\rm{DOS}}}^{*}$ at $\Gamma$-valley \cite{PhysRevB.79.245201}. Because of this consideration, the compressive strain should not be employed since it tends to pull CB at $L$-valley and $X$-valley closer in energy \cite{PhysRevB.79.245201}, resulting in more electrons occupying $X$-valley
, which increases the complexity of the carrier distribution and optical transition. We therefore will focus on tensile strain only.




Effect of uniaxial and biaxial tensile strain on $E_{1}$ and $E_{1}+\Delta_{1}$ optical transitions along [100], [110] and [111] crystal orientation is illustrated Fig. \ref{fig2} (a)-(f).
In comparison with the relaxed situation, the strain may or may not lift the degeneracy of the four $L$-valleys depending on the direction in which it is applied. Similar phenomena around $X$-valley
have been observed in Si \cite{rideau2006strained}. For the convenience of discussion, we shall label the four $L$-valleys as $L_{1}$ ([111]), $L_{2}$ ([11$\overline{1}$]), $L_{3}$ ([1$\overline{1}$1]) and $L_{4}$ ([$\overline{1}$11]).
For the tensile strain along [100], the degeneracy in both CB and VB is not broken and the energy position of the four $L$-valleys are shifted the same amount \cite{rideau2006strained, 8871170, ESCALANTE2018223} as shown in Fig. \ref{fig2} (a) and (d) while the VB separation between HH and LH is widened. On the contrary, the tensile strain along [110] and [111] both break the degeneracy of $L$-valleys as well as the HH and LH VB, dividing the four $L$-valleys into two groups: one having $L_{1,2}$ and $L_{3,4}$ as shown in Fig. \ref{fig2} (b) and (e), and the other $L_{1}$ and $L_{2,3,4}$ in Fig. \ref{fig2} (c) and (f), respectively. The relative energy positions of the $L$-valleys depend on the strength and type of tensile strain, namely, uniaxial or biaxial.
The results of the energy dependence of $E_{1}$ and $E_{1}+\Delta_{1}$ optical transitions on the strength of uniaxial and biaxial tensile strain along different directions are shown in Fig. \ref{fig2} (g)-(l). These variations induced by the tensile strain directly influence the profile of ODF and ODFSD as the detailed analysis will show below according to a two-dimension model \cite{Xuchi1}.

Different from the relaxed Ge situation, because the tensile strain may break the degeneracy of the four $L$-valleys, it is necessary to separately calculate the contribution individually from each valley since they may each have different electron population, and then sum them up to obtain the total ODF. The imaginary part of the total ODF $\varepsilon_2\left(E\right)$ from the four $L$-valleys in the presence of tensile strain could be expressed as
\begin{widetext}
\begin{equation}\label{Eq1}
\varepsilon_{2}(E)=\sum_{L_{1,2,3,4}}\frac{2 e^{2} \bar{P}^{2} \mu_{\perp}}{3 m^{2} E^{2}}H(E-E_{1})\int_{-k_{\max }} ^{k_{\max }} d k_{z}\left\{1-f\left[E_{c}\left(E, k_{z}^{2}\right)\right]\right\},
\end{equation}
\end{widetext}
where the ${\bar{P}}^2$ is the square of the average momentum matrix element, $\mu_\bot$ is the transverse reduced electron-hole mass and $k_{\mathrm{max}}$ is the length of the parallel band along $L-\Gamma$.
$H\left(E-E_1\right)$ is the Heaviside step function and $f\left(E\right)$ is the Fermi function whose argument $E_{c}$ can be written as
\begin{equation}\label{Eq2}
 E_{c}\left(E, k_{z}^{2}\right)=\frac{\hbar^{2} k_{z}^{2}}{2 m_{\|}}+\left(E-E_{1}\right) \frac{\mu_{\perp}}{m_{\perp}}
\end{equation}
where $m_{||}$ and $m_\bot$ are the longitudinal and transverse effective mass, respectively. Having calculated the imaginary part of ODF, its real part $\varepsilon_1\left(E\right)$ could be obtained utilizing the Kramers-Kronig relationship. Here we assume that ${\bar{P}}^2$ is unchanged under tensile strain with the same value given in \cite{Xuchi1} and Fermi level is fiexed to be 0.13 eV above the CBM. All the parameters needed for the calculation of ODF and ODFSD such as the energy positions of the four $L$-valleys, optical transition energies, longitude and
transverse effective masses \cite{Xuchi1} can all be obtained from the 30-band $k \cdot p$ model with strain \cite{rideau2006strained,Zhigang,8871170} and the temperature is fixed at 77K.

\begin{figure}[htb]
\includegraphics[width=3.4in]{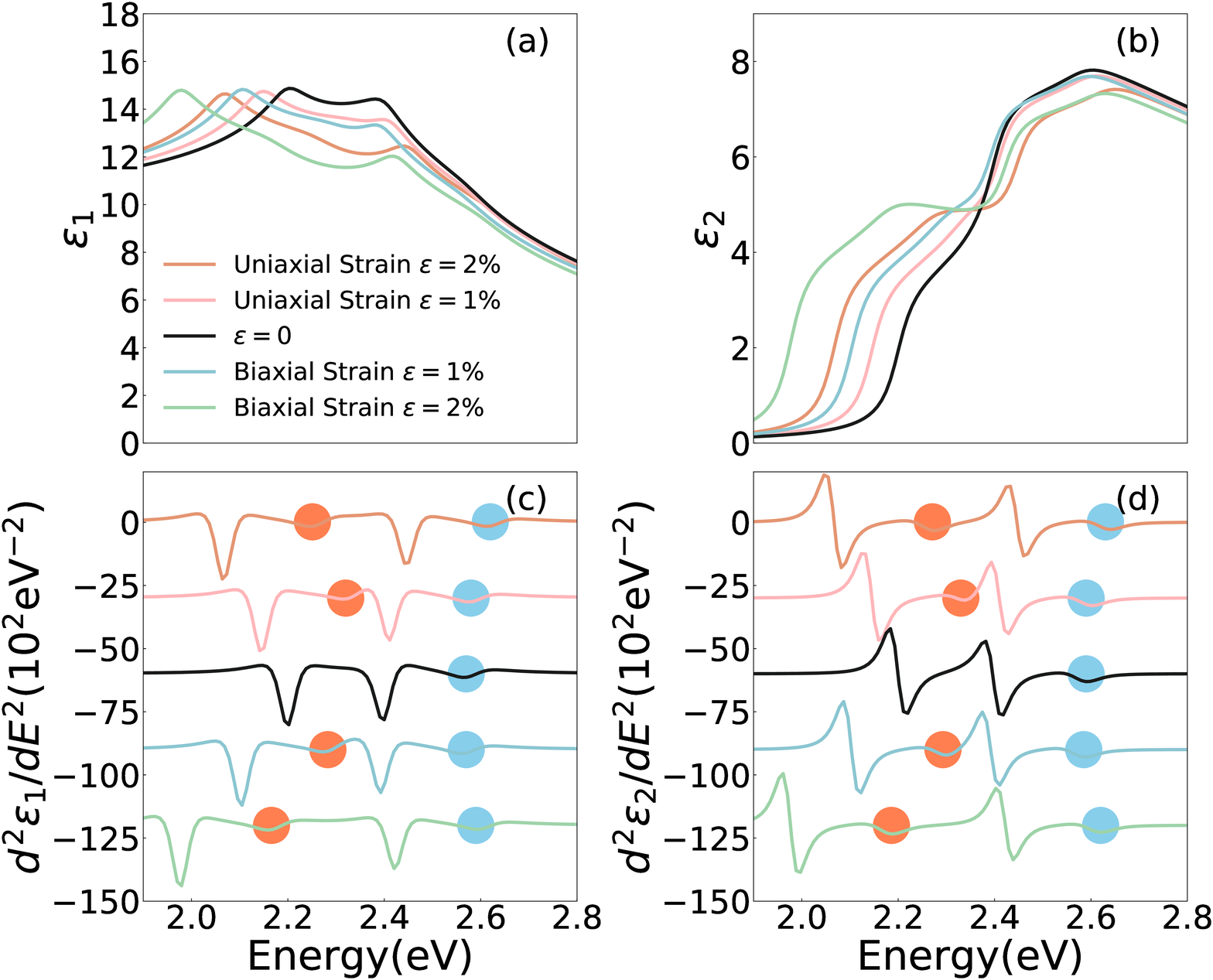}
\caption{The top two panels [(a) and (b)] are the real and imaginary parts of the calculated ODF and the bottom two panels [(c) and (d)] are those of the ODFSD of doped $n$-type Ge near $E_{1}$ and $E_{1}+\Delta_{1}$ gaps under tensile strain along [100]. The black curves denote the relaxed situation and other curves in color represent the strained situation.
Compared with the unstrained Ge where only the PFS associated with the $E_{1}+\Delta_{1}$ optical transition (light blue circles) in (c) and (d) can be observed, tensile strain along [100] of either uniaxial or biaxial makes PFS of both $E_{1}$ (orange circles) and $E_{1}+\Delta_{1}$ (light-blue circles) observable simultaneously as shown in (c) and (d).}
\label{fig3}
\end{figure}
The calculated ODF and ODFSD according to the Eq. \ref{Eq1} in the presence of uniaxial and biaxial tensile strain along [100] are shown in Fig. \ref{fig3} (a)-(d) where the black curves denote the situation of relaxed Ge as shown in \cite{Xuchi1} and other curves in color represent the  situations with strain.
Since the ODFSD exhibits stronger spectral feature than ODF as shown in Fig. \ref{fig3}, we shall focus on the discussion of ODFSD in the presence of the tensile strain.
\begin{figure}[htb]
\includegraphics[width=3.4in]{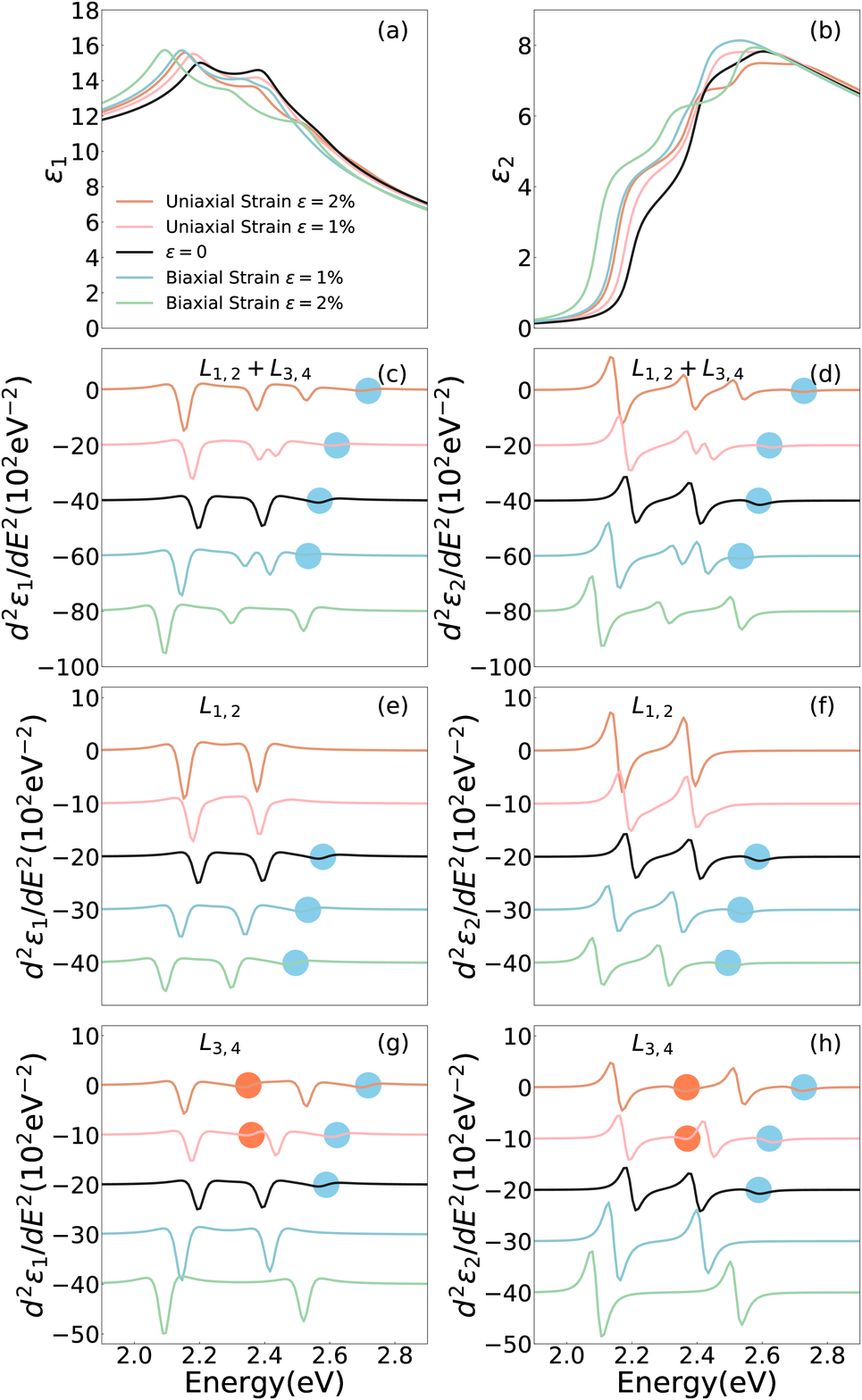}
\caption{(a)-(d) are the real and imaginary part of the calculated ODF and ODFSD under either uniaxial or biaxial tensile strain along [110]. (e)-(h) are the corresponding contributions from $L_{1,2}$-valleys and $L_{3,4}$-valleys under either uniaxial or biaxial strain along [110]. The orange and light-blue circles donate the PFS associated with $E_{1}$ and $E_{1}+\Delta_{1}$ optical transitions, respectively.}
\label{fig4}
\end{figure}
We can see from Fig. \ref{fig3} (c) and (d) that both the uniaxial and biaxial strain induce the red-shift of PFS associated with $E_{1}$ optical transition labeled by the orange circles in real and imaginary part of the ODFSD, but blue-shift for the PFS associated with $E_{1}+\Delta_{1}$ optical transition as labeled by light blue circles. These are consistent with what can be seen in Fig. \ref{fig3} (g) and (j) that the energy of $E_{1}$
optical transition decreases with the increase of tensile strain while that of $E_{1}+\Delta_{1}$ increases slightly. It should be noted that the increase of $\Delta_{1}$ under the tensile strain also contributes to the separation of $E_{1}$ and $E_{1}+\Delta_{1}$ optical transitions. The widening separation between the $E_{1}$ and $E_{1}+\Delta_{1}$ optical transition allows for the PFS of $E_{1}$ optical transition can be more clearly isolated, making it observable for in experimental measurement.

Because of the large DOS effective mass in the $L$-valleys $m_{L_{\rm{DOS}}}^{*}$, heavy doping of $6\times10^{19}$ cm$^{-3}$, is required to raise the Fermi level 0.13 eV above the CBM in relaxed Ge, presenting difficulty in experiment to observe the PFS features. The lifting of the $L$-valley degeneracy offers the possibility of reducing the doping required to observe the PFS. As seen in Fig. \ref{fig2} (b) and (e), both uniaxial and biaxial strain along [110] split the degenerate energy level of the four $L$-valleys into two, $L_{1,2}$ and $L_{3,4}$, each consisting of two $L$-valleys, as a result, roughly half of the electrons are needed to populate the lower $L$-valleys for PFS observation. In the case of tensile strain applied along [111], the degeneracy is lifted in such a way that either one of the four $L$-valleys sits below the other three as shown in Fig. \ref{fig2} (f), or the other way around as shown in Fig. \ref{fig2} (c), offering the possibility of lowering the doping to either a quarter or three quarters of the doping needed. We shall look at the observability of the PFS for each of these cases below.

When the tensile strain along [110] is introduced, the degeneracy lifting of the four $L$-valleys leads to the split of both $E_{1}$ and $E_{1}+\Delta_{1}$ optical transitions to $L_{1,2}$-valleys and $L_{3,4}$-valleys as shown in Fig. \ref{fig2} (b) and (e), dramatically altering the spectral profiles of the ODF and ODFSD as shown in Fig. \ref{fig4} (a)-(d). Different from the uniaxial and biaxial tensile strain along [100] that only widens the separation of $E_{1}$ and $E_{1}+\Delta_{1}$ optical transitions, the uniaxial and biaxial tensile strain along [110] actually induces rather distinct behavior in PFS because of the different occupation populations of the doped electrons at $L_{1,2}$-valleys and $L_{3,4}$-valleys as the result of their degeneracy lifting. To maintain the Fermi level at the same 0.13 eV above the CBM, only nearly half of the doped electrons are required because only two out of the four $L$-valleys need to be populated. Since the total ODF and ODFSD are the sum of contributions from the two inequivalent groups: $L_{1,2}$-valleys and $L_{3,4}$-valleys, we need to calculate them individually. Under the uniaxial tensile strain along [110], the CBM at $L_{3,4}$-valleys is lower than that of $L_{1,2}$-valleys as shown in Fig. \ref{fig2} (b) and as a result, almost all of the doped electrons occupy $L_{3,4}$-valleys with a small fraction in $L_{1,2}$-valleys. The profile of ODFSD from
$L_{1,2}$-valleys therefore behaves more like undoped intrinsic Ge as shown in Fig. \ref{fig4} (e) and (f) while that from $L_{3,4}$-valleys behave like the heavily doped Ge as shown in Fig. \ref{fig4} (g) and (h) for the real and imaginary part of the ODFSD, respectively. Under the biaxial tensile strain along [110], however, the opposite occurred.
\begin{figure}[htb]
\includegraphics[width=3.4in]{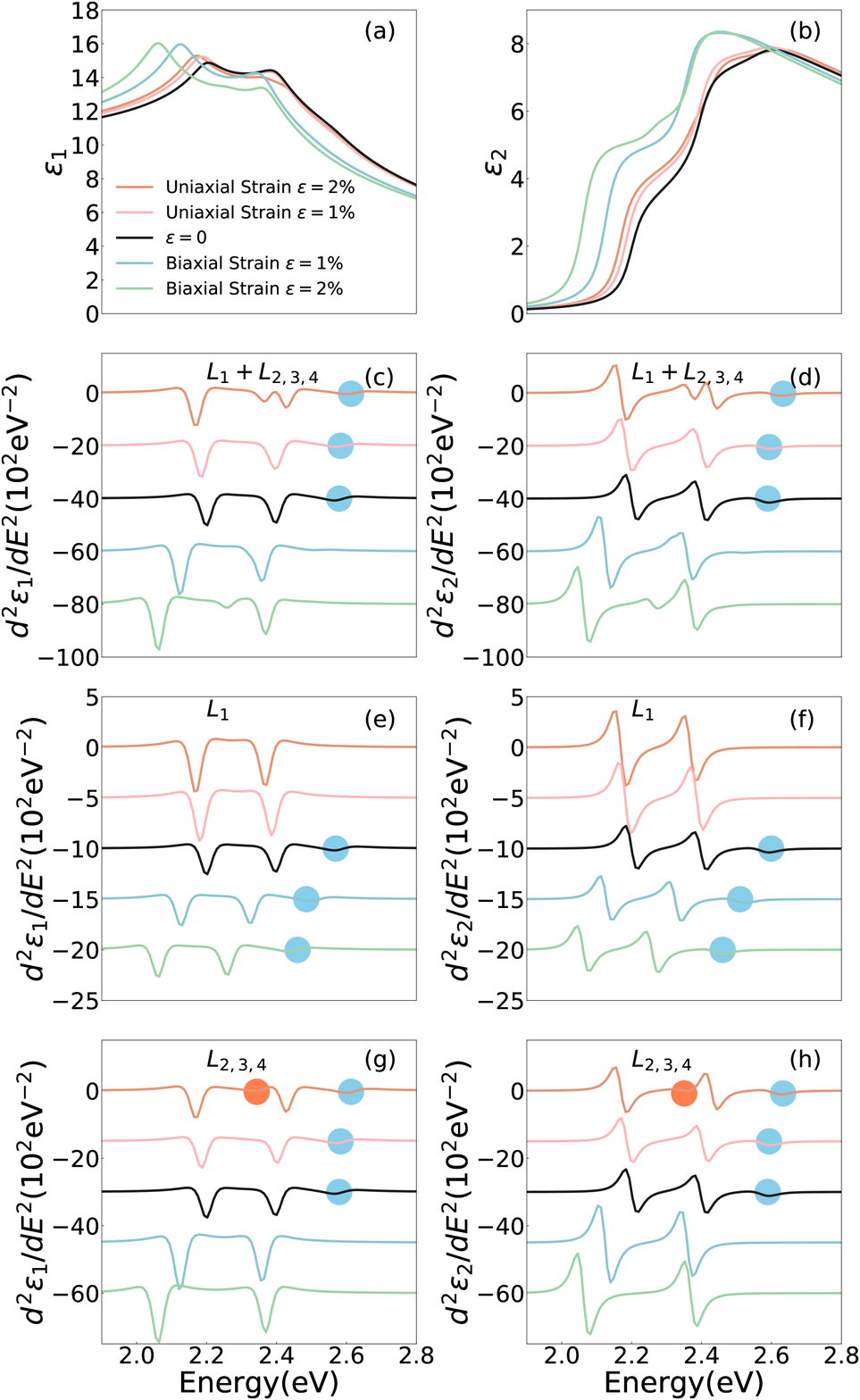}
\caption{FIG. 5. (a)-(d) are the real and imaginary part of the calculated ODF and ODFSD under either uniaxial or biaxial tensile strain along [111]. (e)-(h) are the corresponding contributions from $L_{1}$ and $L_{2,3,4}$-valleys under either uniaxial or biaxial strain along [111]. The orange and light-blue circles donate the PFS associated with $E_{1}$ and $E_{1}+\Delta_{1}$ optical transitions, respectively. }
\label{fig5}
\end{figure}
The total real and imaginary part of the ODFSD shown in Fig. \ref{fig4} (c) and (d) can be obtained as the sum of Fig. \ref{fig4} (e) and (f), and Fig. \ref{fig4} (g) and (h), respectively. It can be seen from Fig. \ref{fig4} (c) and (d) that with the tensile strain along [110] direction, while it reduces the doping requirement to half, the PFS feature associated with $E_{1}$ transition that appeared in the uniaxial situation shown in Fig. \ref{fig4} (g) and (h) disappears in the total ODFSD as they sit too close to the ordinary $E_{1}+\Delta_{1}$ optical transition. Some new features, however, can be observed in comparison with the relaxed Ge with heavy doping, including more dips and peaks resulting from the degeneracy lifting of the four $L$-valleys into $L_{1,2}$-valleys and $L_{3,4}$-valleys, allowing for additional $E_{1}$ and $E_{1}+\Delta_{1}$ transitions as shown in Fig \ref{fig2} (b) and (e). The PFS associated with $E_{1}+\Delta_{1}$ optical transition disappears as shown in Fig. \ref{fig4} (c) and (d) when the strength of the biaxial strain exceeds a certain value, say $2\%$, because of its overlap with the ordinary $E_{1}+\Delta_{1}$ optical transition for $L_{3,4}$-valleys.

For the [111] tensile strain, the four $L$-valleys are split into two groups consisting of $L_{1}$ and $L_{2,3,4}$-valleys. The relative energy positions of the two groups depend on the type and strength of tensile strain as shown in Fig. \ref{fig2} (c) and (f). The real and imaginary part of the ODF and ODFSD are shown in Fig. \ref{fig5} (a)-(d). For uniaxial strain, $L_{2,3,4}$-valleys sit lower than $L_{1}$, therefore, three quarters of the doping required for relaxed Ge are needed to observe the PFS and contribution comes from the $L_{2,3,4}$-valleys as shown in Fig \ref{fig5} (g) and (h). For biaxial strain, however, the situation is reversed, while the contribution comes from $L_{1}$-valley for which only a quarter of doping is required, but, PFS features in the total ODF and ODFSD are unobservable. But similar to the tensile strain along [110], additional features appear due to the split of $E_{1}$ and $E_{1}+\Delta_{1}$ transitions.

In summary, we systematically study the PSF features in ODF and ODFSD in heavily doped Ge under the uniaxial and biaxial tensile strain along [100], [110] and [111]. Our investigation theory predicts that the uniaxial and biaxial tensile strain along [100] direction can lead to experimental observation of PFS associated $E_{1}$ optical transition that is previously unresolvable in the heavily doped relaxed Ge. The tensile strain applied along [110] and [111] can lower the doping required for the observation of PFS associated $E_{1}+\Delta_{1}$ optical transition. In the case of [110] tensile strain, the doping can be reduced to half. In the case of [111] strain, the doping can only be to about three quarters. This study indicates that tensile strain is an effective tool in eliminating the difficulty in separating the $E_{1}$ and $E_{1}+\Delta_{1}$ optical transitions in the relaxed Ge and in revealing PFS features in ODF and ODFSD.

Z. Song acknowledges the helpful discussion with José Menéndez and Chi Xu in regard to the calculation of PFS and the experimental measurement. WJ Fan acknowledges the funding support (NRF-CRP19-2017-01). The computation of this work was partially performed on resources of the National Supercomputing Centre, Singapore. G Sun acknowledges the grant support (FA9550-19-1-1034) from the Air Force Office of Scientific Research.

\bibliography{Ge_filling}

\end{document}